\documentclass[aps,showpacs,amsmath,amssymb,twocolumn,pra,superscriptaddress,notitlepage]{revtex4-2}

\usepackage{qcircuit}
\usepackage{amsmath,bm}
\usepackage[dvips]{graphicx}
\usepackage{amsmath,amssymb,amsthm,mathrsfs,amsfonts,dsfont}
\usepackage{subfigure, epsfig}
\usepackage{braket}
\usepackage{bm}
\usepackage{enumerate}
\usepackage{color}
\usepackage{graphicx}
\usepackage{algorithm}
\usepackage{algorithmic}
\usepackage{braket}
\usepackage{comment}
\usepackage{appendix}
\usepackage{here}
\usepackage{tabularx}
\usepackage{physics}
\usepackage{mathtools}
\usepackage{multirow} 
\usepackage{float}

\begin{document}

\title{Adaptive measurement strategy for quantum subspace methods}

\author{Yuma Nakamura}
\email{yuma.nakamura1@ibm.com}
\affiliation{Healthcare \textnormal{\&} Life Science, IBM Japan, 19-21 Nihonbashi Hakozaki-cho, Chuo-ku, Tokyo 103-8510, Japan}

\author{Yoshichika Yano}
\affiliation{Department of Applied Physics, University of Tokyo,
7-3-1 Hongo, Bunkyo-ku, Tokyo 113-8656, Japan}

\author{Nobuyuki Yoshioka}
\email{nyoshioka@ap.t.u-tokyo.ac.jp}
\affiliation{Department of Applied Physics, University of Tokyo,
7-3-1 Hongo, Bunkyo-ku, Tokyo 113-8656, Japan}
\affiliation{Theoretical Quantum Physics Laboratory, RIKEN Cluster for Pioneering Research (CPR), Wako-shi, Saitama 351-0198, Japan}
\affiliation{JST, PRESTO, 4-1-8 Honcho, Kawaguchi, Saitama 332-0012, Japan}

\begin{abstract}
Estimation of physical observables for unknown quantum states is an important problem that underlies a wide range of fields, including quantum information processing, quantum physics, and quantum chemistry.
In the context of quantum computation, in particular, existing studies have mainly focused on holistic state tomography or estimation on specific observables with known classical descriptions, while this lacks the important class of problems where the estimation target itself relies on the measurement outcome.
In this work, we propose an adaptive measurement optimization method that is useful for the quantum subspace methods, namely the variational simulation methods that utilize classical postprocessing on measurement outcomes.
The proposed method first determines the measurement protocol for classically simulatable states, and then adaptively updates the protocol of quantum subspace expansion (QSE) according to the quantum measurement result.
As a numerical demonstration, we have shown for excited-state simulation of molecules that (i) we are able to reduce the number of measurements by an order of magnitude by constructing an appropriate measurement strategy (ii) the adaptive iteration converges successfully even for a strongly correlated molecule of H$_4$.
Our work reveals that the potential of the QSE method can be empowered by elaborated measurement protocols, and opens a path to further pursue efficient quantum measurement techniques in practical computations. 
\end{abstract}

\maketitle
\section{introduction}
\label{section: introduction}
Estimating observables of quantum states is an exceedingly important process that serves as the foundation across a wide range of fields involving quantum technology.
This importance is especially pronounced in the realm of quantum computing~\cite{Feynman1982,nielsen2002quantum} that aims to perform, e.g.,  quantum many-body simulation~\cite{abrams1997, Abrams99, sethuniversal}, solving sparse linear equations~\cite{harrow2009quantum}, and quantum machine learning~\cite{schuld2015introduction, biamonte2017quantum, schuld2021machine}.
Consequently, one of the highest priorities is the development of a measurement protocol that imposes minimal burden on the quantum device.

When one has access to a quantum computer with fault-tolerance such that long coherence is ensured, one may rely on the amplitude estimation that achieves the Heisenberg limit of complexity $O(1/\epsilon)$ in terms of additive error $\epsilon$~\cite{brassard2002quantum}.
There have been some attempts to extend the scheme to a multiple observable estimation regime~\cite{higgins2009demonstrating, huggins2022nearly}, while it is still open how to achieve the optimal query complexity.
Meanwhile, when the coherence of the quantum computer is limited, we shall rather depend on projective measurements with circuit executions over $O(1/\epsilon^2)$ times, either in a near-term device~\cite{peruzzo2014variational, bauer2020quantum, cerezo2021variational} or in early fault-tolerant regime~\cite{lin2022heisenberg, ding2023even}.
Existing works in the context of variational quantum simulations~\cite{peruzzo2014variational, bauer2020quantum, cerezo2021variational} have exploited the classical description of the target observable to be measured: the $L^1$ norm of the Pauli coefficients for the target observable~\cite{kandala_hardware_2017, mcclean_theory_2016}, qubit-wise commutativity of Pauli operators embedded into Minimum Clique Cover (MCC) problem~\cite{bravyi2017tapering, verteletskyi_measurement_2020, Crawford2021efficientquantum, hamamura2020efficient, hillmich2021decision}, and expression of uncertainty for adaptive modification of measurement resource~\cite{garcia2021learning}.
Another important direction is to utilize randomized measurements, which are designed to achieve average-case optimality so that
one may choose target observables arbitrarily after the measurement is done~\cite{enk2012measuring, elben2018renyi, elben2019statistical,huang_predicting_2020,  elben2023randomized}.
For instance, the protocol referred to as the classical shadow (CS) tomography achieves estimation accuracy $\epsilon$ for $M$ observables with $O(\log(M)/\epsilon^2)$ measurements, with prefactors that crucially rely on the set of available positive operator-valued measurements~\cite{huang_predicting_2020, zhao_fermionic_2021, low2022classical, elben2023randomized}. 
When one is interested in some specific observables such as quantum many-body Hamiltonians, one may introduce appropriate bias on the measurement strategy so that the estimation variance is suppressed~\cite{hadfield_measurements_2022, huang2021efficient, nakaji2023measurement}.

We emphasize that existing methods assume two ultimate situations. Namely, it has been assumed that the classical description of the measurement target, e.g., the coefficients to express the target observable as a sum of Pauli operators, is completely known beforehand, or it is completely provided after the measurement.
Thus, we are lacking the solution to the important class of problems where the target observables explicitly rely on the measurement outcome. More generally, this problem is common to tasks where the postprocessing operation depends explicitly on the target quantum state.
This implies that we are not fully utilizing the capacity of quantum subspace methods~\cite{mcclean_hybrid_2017, mcclean2020decoding, yoshioka_gse_2022, yang2023dual, cortes2022quantum, cohn2021quantum, seki2021quantum} or quantum algorithms that utilize error mitigation techniques such as the symmetry expansion~\cite{Cai2021quantumerror, endo2022quantum, tsubouchi2023virtual}, and furthermore early fault-tolerant quantum computers with quantum error detection methods~\cite{PhysRevA.56.33,  PhysRevLett.77.198, PhysRevLett.77.793, Devitt_2013}.
Considering that such post-processing techniques and error mitigation methods are envisioned to remain crucial for both near-term and fault-tolerant quantum devices~\cite{cai2022quantum}, it is critical to overcome the above issue to
 fully harness their computational powers.
 
\begin{figure*}[ht]
\begin{center}
\includegraphics[width=18cm]{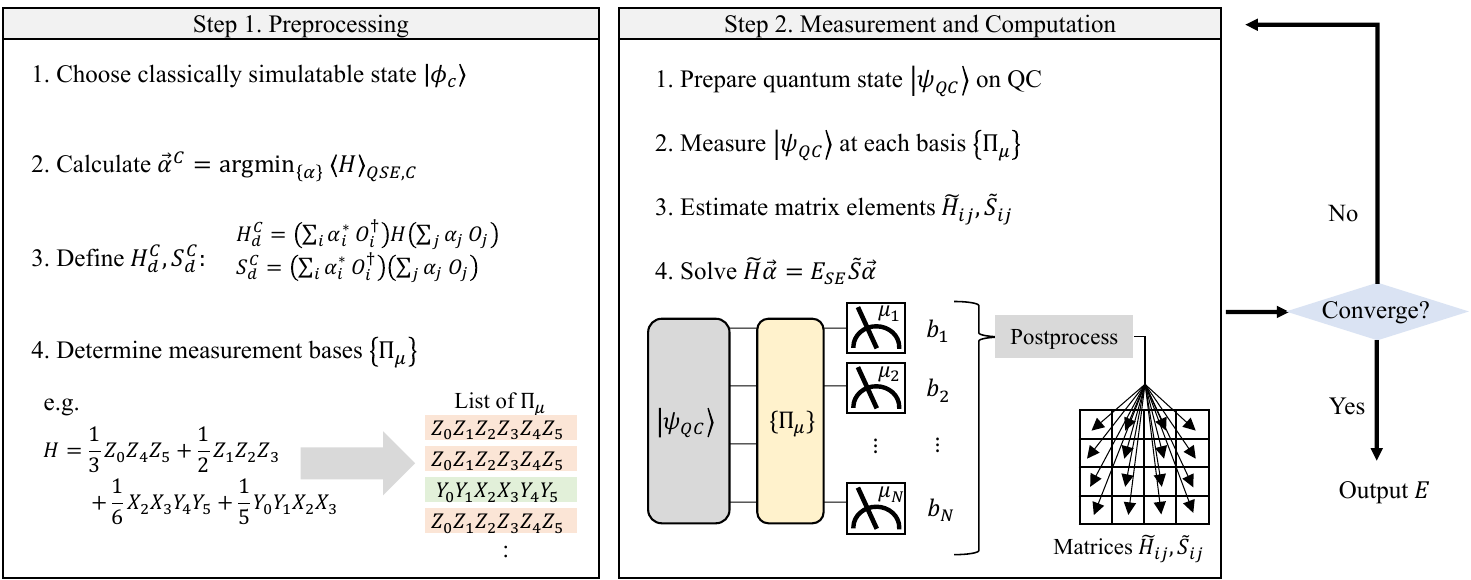}

\end{center}
\vspace*{-0.25cm}
\caption{\label{fig:protocol}
Flow of adaptive measurement strategy construction.
\vspace*{-0.25cm}
}

\end{figure*}

In this study, we propose an adaptive measurement strategy for quantum subspace methods. 
In this protocol, efficient measurement is achieved by adaptively updating the measurement strategy based on post-processed calculation results.
As numerical validation of the proposed scheme, we perform excited state simulation for molecules such as H$_2$, H$_4$, and LiH. 
As a result, we achieve a significant reduction in the number of measurements, approximately 3 to 10 times fewer compared to a naive measurement approach. 
We further show that the adaptive iteration converges rapidly even for the strongly correlated regime in H$_4$, supporting the effectiveness of the proposed method.

\section{METHODS}
\subsection{Quantum subspace expansion}
\label{sec_1}
Quantum subspace methods are a collection of techniques in which the many-body Hamiltonian on the entire Hilbert space is projected on one of the appropriate subspaces by applying classical post-processing to information obtained through measurements on a quantum device which is mainly a quantum computer in practice~\cite{mcclean_hybrid_2017, mcclean2020decoding,colless2018computation, dhawan2023quantum, cohn2021quantum, gao2021applications, yoshioka_gse_2022, yoshioka_periodic_2022, yang2023dual, ohkura2023leveraging}.
After the theoretical proposal by McClean {\it et al.} to utilize the near-term quantum device for excited-state calculation~\cite{mcclean_hybrid_2017}, numerous works have exploited the similarity between the classical power methods to enhance the capability of the quantum subspace method~\cite{cohn2021quantum, seki2021quantum}.
Important extensions include many-body calculations such as real-time evolution~\cite{heya2023subspace}, finite-temperature calculations~\cite{pokhilko2022iterative}, non-equilibrium steady state calculations~\cite{yoshioka_ness_2020}, and imaginary time Green's functions~\cite{dhawan2023quantum}.

Such theoretical advancements have also accelerated the experimental demonstrations in noisy quantum devices.
For instance, Colless {\it et al}.~\cite{colless2018computation} experimentally evaluated the excited-state energies of the H$_2$ molecule within the subspace spanned by single Pauli operators $ \{O_j \} = \{\sigma_\alpha^k \ | \  \alpha \in \{x, y, z\}, k \in \{1,2\}\} $ using the ground state as the reference state, in which the effect of error mitigation was also observed.
As a variant of the subspace method, Gao {\it et al.}~\cite{gao2021applications} simulated the equation-of-motion Hamiltonian for organic diodes.
For a more exhaustive review of both theoretical and experimental works, we guide the readers to Ref.~\cite{motta2023subspace}.

Now let us describe the formalism of the subspace methods.
In the following, we specifically consider the computation of eigenvalues of a Hamiltonian $H$ with $N$ qubits.
The simulation of eigenstates is often discriminated from others explicitly by using the term Quantum Subspace Expansion (QSE), and thus hereafter we follow the convention.
With a set of $k$ reference states $\{\ket{\psi}_k\}_k$ realized on a quantum computer, let us assume that we have a set of operators $\{O_{j}^{(k)}\}_{j, k}$ which yields a subspace $\mathcal{S} = {\rm Span}\{O_{j}^{(k)} \ket{\psi_k}\}$, whose bases are typically non-orthogonal to each other. For the purpose of eigenstate calculation, we define the following variational state: 
\begin{eqnarray}
    \ket{\psi} = \sum_{j,k} \alpha_j O_j^{(k)} \ket{\psi_k} \label{eq:qse_psi},
\end{eqnarray}
where the coefficients $\{\alpha_j^{(k)}\}_{j, k}$ are variational parameters that are determined based on the desired computational target. When the target is the energy eigenstate, the Rayleigh-Ritz variational condition leads to the following generalized eigenvalue equation:
\begin{eqnarray}
    \widetilde{H}\vec{\alpha} = E_{\rm SE} \widetilde{S} \vec{\alpha}, \label{eq:qse}
\end{eqnarray}
where $E_{\rm SE}$ represents the eigenvalues, and $\widetilde{H}$ and $\widetilde{S}$ are matrices representing the Hamiltonian and overlap between the bases, respectively, restricted to the subspace. The matrix elements are given by the following equation:
\begin{eqnarray}
    \widetilde{H}_{ij}^{(kk')} &=& \mel**{\psi_k}{(O_i^{(k)})^\dag  H O_j^{(k')}}{\psi_k'}, \label{eq:h_mat}\\
    \widetilde{S}_{ij}^{(kk')} &=& \mel**{\psi_k}{(O_i^{(k)})^\dag O_j^{(k')}}{\psi_{k'}}. \label{eq:s_mat}
\end{eqnarray}

Hereafter, for simplicity, we consider a situation where there is only one reference state given as $\ket{\psi_{\rm QC}}$ that is assumed to be realized on a quantum computer. 
In this case, the final energy $E_{\rm SE}$ to be determined is expressed as follows:
\begin{eqnarray}
    E_{\rm SE} &=& \frac{\vec{\alpha}^\dag \widetilde{H}\vec{\alpha}}{\vec{\alpha}^\dag \widetilde{S}\vec{\alpha}} = \frac{\sum_{ij}\alpha_i^* \widetilde{H}_{ij} \alpha_j}{\sum_{ij}\alpha_i^* \widetilde{S}_{ij} \alpha_j} \\
    &=& \frac{\sum_{ij}\sum_Q \alpha_i^* \alpha_j c_Q \ev*{O_i^\dag P_Q O_j}{\psi_{QC}}}{\sum_{ij} \alpha_i^* \alpha_j \ev*{O_i^\dag  O_j}{\psi_{QC}}},
    \label{eq:qse_energy}
\end{eqnarray}
where the Hamiltonian is expressed as $H = \sum_Q c_{\rm Q} P_Q~(c_Q\in\mathbb{R}, P_Q\in\{I, X, Y, Z\}^{\otimes N})$ in the second line.

To solve the generalized eigenvalue equation in Eq.~\eqref{eq:qse}, quantum measurements are necessary to determine the matrix elements $\widetilde{H}_{ij}$ and $\widetilde{S}_{ij}$ as defined by equations \eqref{eq:h_mat} and~\eqref{eq:s_mat}. 
To make efficient use of the limited measurement resources, it is desirable to obtain precise results with as few measurements as possible.
However, the optimal measurement strategy for QSE cannot be determined solely from the Hamiltonian; it depends on the solution of the generalized eigenvalue equation itself. 
While the error of $E_{\rm SE}$ due to the measurement error can be upper-bounded~\cite{yoshioka_gse_2022}, it is favorable to minimize the error by choosing a suboptimal measurement strategy.
Therefore, here we develop adaptive computation methods that are inspired by solvers for nonlinear eigenvalue problems.

\subsection{Adaptive measurement strategy for QSE }

First, it should be noted that Eq.~\eqref{eq:qse_energy} is equivalent to the estimation of the expectation values of dressed operators as follows:
\begin{eqnarray}
    H_d &=& \sum_{ij} \sum_Q \alpha_i^* \alpha_j c_Q O_i^\dag P_Q O_j, \label{eq:h_op}\\
    S_d &=& \sum_{ij} \alpha_i^* \alpha_j  O_i^\dag O_j\label{eq:s_op}.
\end{eqnarray}
The expectation values of these operators can be understood to provide energy as $E_{\rm SE} = \ev{H_d}/\ev{S_d}$, where $\ev*{\cdot}:=\ev{\cdot}{\psi_{\rm QC}}$.
As mentioned earlier, once one has performed the quantum measurement and all the matrix elements of $\tilde{H}$ and $\tilde{S}$ are known, we can apply known measurement strategies such as MCC methods. However, in practice, these expressions are unknown before quantum measurements. This motivates us to propose an adaptive procedure as illustrated in 
Fig.~\ref{fig:protocol} (See the codes available via GitHub~\cite{shadowTools} for details.)

\noindent\underbar{Step 1: Preprocessing in classical computation}
\begin{itemize}
    \item[1.] Given the Hamiltonian $H$ and a set of excitation operators $\{O_i\}$, select a state $\ket{\phi_{\rm C}}$ that can be simulated classically.
    \item[2.] Solve the generalized eigenvalue equation~\eqref{eq:qse} based on matrices $\tilde{H}$  and $\tilde{S}$ computed solely on a classical computer.
    This yields a set of the coefficients $\{\alpha^{\rm C}_i\}_i$.
    \item[3.] Using the obtained coefficients mentioned above, calculate $H_d^{(C)}$ and $S_d^{(C)}$ as in Eq.~\eqref{eq:h_op} and~\eqref{eq:s_op}.
    \item[4.] Apply measurement basis optimization methods, such as the Locally-Biased Classical Shadow (LBCS)~\cite{hadfield_measurements_2022} or MCC~\cite{verteletskyi_measurement_2020}, to $H_d^{(C)}$ and $S_d^{(C)}$. This allows us to determine the measurement strategy, i.e., the distribution of the measurement bases $\{\Pi^{(m=1)}_\mu\}$, where $m$ and $\mu=(\mu_1, ..., \mu_N)~(\mu_k\in \{X, Y, Z\})$ denote the iteration index and the set of single-qubit measurement bases, respectively. One may set a maximum number of bases in practice.
\end{itemize}

\noindent\underbar{Step 2: Quantum measurement and computation}\\
    Repeat the following.
\begin{itemize}
        \item[1.] Prepare the reference quantum state $\ket{\psi_{\rm QC}}$ on the quantum computer.
        \item[2.] Perform quantum measurements in the Pauli basis based on the distribution $\{\Pi^{(m)}_\mu\}_\mu$ with a given measurement budget. 
        \item[3.]
        Obtain $\widetilde{H}^{(m)}$ and $\widetilde{S}^{(m)}$ under appropriate regularization (See Appendix~\ref{sec:reg} for details). 
        \item[4.] Solve the generalized eigenvalue equation to obtain the energy $E^{(m)}$ and the set of coefficients $\{\alpha_i^{(m)}\}_i$.
        \item[5.] If the solution has converged or the measurement budget has been all consumed, terminate the computation. If not, update the dressed operator $H_d$ and $S_d$ based on the latest set of coefficients $\{\alpha_i^{(m)}\}_i$,
        create a distribution of measurement bases $\{\Pi_\mu^{(m+1)}\}_\mu$ using coefficients $\{\alpha^{(m)}\}$, and proceed to the $(m+1)$-th round of measurement and computation.
\end{itemize}

\begin{table*}[hbtp]
    \centering
    \renewcommand{\arraystretch}{1.3}
    \begin{tabular}{c|c|cc|cc}
        \hline
        \multirow{2}{*}{Molecules} & Total  & \multicolumn{2}{c|}{Equally-Distributed} & \multicolumn{2}{c}{Optimization Subroutine}  \\
            & Measurements & $|E_{\rm est} - E_{\rm exact}|$ &Std($E_{\rm est}$) & \multicolumn{1}{l}{$|E_{\rm est} - E_{\rm exact}|$} & Std($E_{\rm est}$) \\ \hline
        H$_2$, 4 qubits & $10^3$ & 0.0226 & 0.0242 & $\leq 10^{-5}$ & $\leq 10^{-5}$ \\ \hline
        \multirow{4}{*}{H$_2$, 8 qubits} & $10^3$ & 0.2167 & 0.2285 & 0.1267 & 0.1254 \\
            & $10^4$ & 0.0777 & 0.1032 & 0.0526 & 0.0546 \\
            & $10^5$ & 0.0197 & 0.0228 & 0.0188 & 0.0172 \\
            & $10^6$ & 0.0089 & 0.0085 & 0.0164 & 0.0111 \\ \hline
        \multirow{4}{*}{LiH, 10 qubits} & $10^3$ & 0.9571 & 0.9973 & 0.0733 & 0.0978 \\
            & $10^4$ & 0.1146 & 0.1166 & 0.0417 & 0.0341 \\
            & $10^5$ & 0.0525 & 0.0449 & 0.0305 & 0.0240 \\
            & $10^6$ & 0.0346 & 0.0373 & 0.0246 & 0.0232 \\ \hline    
        \end{tabular}
    \renewcommand{\arraystretch}{1}
    \caption{
    Comparison between naive equally-distributed measurements and measurement optimization subroutines.
    Note that the former uses LBCS for $\widetilde{H}_{ij}$ (matrix-element-wise LBCS), while the latter applies it to the dressed operator $H_d$.
    Here, we show the mean absolute errors and standard deviations over 10 trials with the total numbers of measurement shots given by $10^3$, $10^4$, $10^5$, and $10^6$. 
    The basis sets for the second-quantized molecular Hamiltonians are taken for H$_2$ as STO-3G (4 qubits) and 6-31G (8 qubits), and for LiH as STO-6G (10 qubits with frozen 1$s$ orbital). 
    }
    \label{table:data_protocol_advantage}
\end{table*}

Three remarks are in order.
First, for the classical preprocessing in Step 1, it is desirable to use a state $\ket{\phi_C}$ that does not completely disentangle qubits.
For example, if we adopt the Hartree-Fock state for a quantum chemistry Hamiltonian, the expectation value of any electron excitation operators becomes zero as, e.g., 
$\langle {\rm HF}| a_i^\dag a_{j(\neq i)} | {\rm HF}\rangle = 0$. 
In this case, it is not possible to construct a non-trivial measurement strategy, since the electron excitation operators are not included in $H_d^{C}$ and $S_d^C$. 
To circumvent this issue, it is favorable to adopt quantum states that incorporate electronic excitations, such as CIS (Configuration Interaction Singles), CISD (Configuration Interaction Singles and Doubles), and others. 
Generally, the requirement for $\ket{\phi_C}$ is that it allows for polynomial-time computation of expectation values over polynomially many Pauli operators, such as the (near) stabilizer states~\cite{gottesman1998heisenberg}, matrix product states~\cite{RevModPhys.77.259, verstraete2008matrix, cirac2021matrix} or (multi) Slater-determinant states~\cite{PhysRevB.55.7464, PhysRevLett.83.2777}. The impact of the overlap between the true solution and $\ket{\phi_C}$ on the final accuracy is an interesting challenge to investigate.

Second, the number of iterations required for convergence in Step 2 is generally unknown when the number of measurements is limited.
As we discuss later in Sec.~\ref{sec:results}, our numerical simulation with shot noise suggests that only a single iteration is sufficient for the case of H$_4$ molecule. 

Third, the convergence criteria can be practically taken as the variance or fluctuation for the output energy. For instance, one may compare the latest and second-latest output energy whether the difference is less than, e.g., the chemical accuracy, while one may also consider the fluctuation of the moving average as well. In our experiment for H$_4$ provided in Sec.~\ref{sec:results}, this is typically achieved in a few steps,  while we have proceeded up to 10 iterations for the sake of demonstration.

\subsection{Measurement basis optimization subroutines}\label{subsec:subroutine}
The proposed method includes a subroutine for optimizing measurement bases  $\{\Pi_\mu^{(m)}\}$ using the classical representation of operators $H_d^{m}$ and $S_d^{m}$. In the case of multiqubit systems, specifically, we can categorize measurement strategies mainly into two groups.
The first group employs randomized measurements with a focus on specific operators, while the second group determines measurement bases based on the commutativity between Pauli operators. 
In this study, we focus on Locally-Biased Classical Shadows (LBCS) ~\cite{hadfield_measurements_2022} and Derandomized Classical Shadows (DCS) ~\cite{huang2021efficient} for the former approach, 
while for the latter, we adopt Minimum Clique Cover (MCC) grouping~\cite{verteletskyi_measurement_2020} and Overlapped Grouping Measurement (OGM) ~\cite{wu_overlappedgrouping_2023}. 
For detailed information on each method, refer to Appendix~\ref{app:subroutine}.

\section{Results} \label{sec:results}
Here we provide numerical demonstrations for the proposed adaptive scheme. First, we discuss and compare the measurement basis optimization subroutines in Sec.~\ref{subsec:subroutine_selection}, and then in Sec.~\ref{subsec:convergence_adaptive_iteration} we show that the adaptive protocol converges successfully within chemical accuracy for a strongly correlated molecule of H$_4$.

Throughout the demonstrations, we exclusively consider the lowest-energy eigenstate within different particle sectors from the ground state energy for the second-quantized molecular Hamiltonian:
\begin{eqnarray}\label{eq:fermionic_ham}
    H=\sum_{pq} t_{pq} a_p^\dag a_q + \sum_{pqrs} v_{pqrs} a_p^\dag a_q^\dag a_r a_s,
\end{eqnarray}
where $a_p^{(\dag)}$ represents the electron annihilation (creation) operators for the $p$-th spin orbital, and $t_{pq}$ and $v_{pqrs}$ represent the amplitudes for one-body and two-body interactions, respectively.
The qubit representation $H = \sum_Q c_Q P_Q$ is obtained via the Jordan-Wigner transformation from the fermionic representation of Eq.~\eqref{eq:fermionic_ham}.
We note that, in order to focus on the statistical error associated with the measurement strategy, we have assumed that the 
exact ground state $\ket{\psi_{\rm QC}} = \ket{\psi_{\rm GS}}$ is realized on a quantum computer.

\subsection{Comparison of Measurement basis optimization subroutines for QSE} \label{subsec:subroutine_selection}
Here, we confirm the significance of the measurement basis optimization subroutine that it is as crucial to the final accuracy as the introduction of the adaptive method itself.
In particular, we consider the following cases: the H$_2$ molecule with the STO-3G basis set ($N=4$) and the 6-31G basis set ($N=8$), 
and the LiH molecule with the STO-6G basis set ($N=10$ under frozen 1$s$ orbital). 
The subspace is spanned as $\mathcal{S}={\rm Span}\{O_i \ket{\psi_{GS}}\}$ with single electron excitations $O_i = a_i$  to simulate the excited states.

First, in Table~\ref{table:data_protocol_advantage}, we compare the accuracy of QSE by the naive measurement strategy 
 and a Pauli-measurement-basis optimization subroutine mentioned in Sec.~\ref{subsec:subroutine}. 
In the naive approach, an equal number of measurements is allocated to each matrix element, and then further distributed among Pauli operators using matrix-element-wise LBCS. 
This results in approximately $O(N_{\rm tot}/N^6)$ measurements for a single Pauli operator under the total measurement resource $N_{\rm tot}$, as each matrix element of $\widetilde{H}$ consists of $O(N^4)$ Pauli operators. 
In contrast, the standard deviation of energy estimations using the LBCS subroutine within the latter approach is below the chemical accuracy for H$_2$~(STO-3G) demonstrating the effectiveness of the measurement basis optimization. Note that the standard deviation is enhanced by around a factor of 3 to 10.
Given that the improvement achieved by simultaneous measurements for the explicit representation of operators was not limited to a constant factor but rather showed polynomial improvement~\cite{gokhale2020n,zhao2020measurement}, we expect that similar improvement shall be present in QSE as well. 

To make a comparison between measurement optimization subroutines, we have computed the mean absolute errors and standard deviations of estimated excited-state energies in CS, LBCS, DCS, MCC and OGM, as shown in Fig.~\ref{fig:qse_result} (see Appendix~\ref{sec:fig_table} for detailed data).
As expected, the measurement optimization methods yield lower estimation variance compared to the CS, which is a simple randomized measurement.
While MCC shows a relatively favorable reduction in error for LiH, it exhibits undesirable overhead for H$_2$.
Nevertheless, both methods follow the scaling by the standard quantum limit of $1/\sqrt{N_{\rm tot}}$.
The performance of LBCS is relatively comparable but superior to CS in terms of both errors and standard deviations.
In a comprehensive assessment, DCS and OGM demonstrate stable performance, with DCS exhibiting a slight advantage within 10 trials.

\begin{figure}[t]
    \begin{center}
    \hspace{-0.75cm}
    \includegraphics[width=22pc,angle=0]{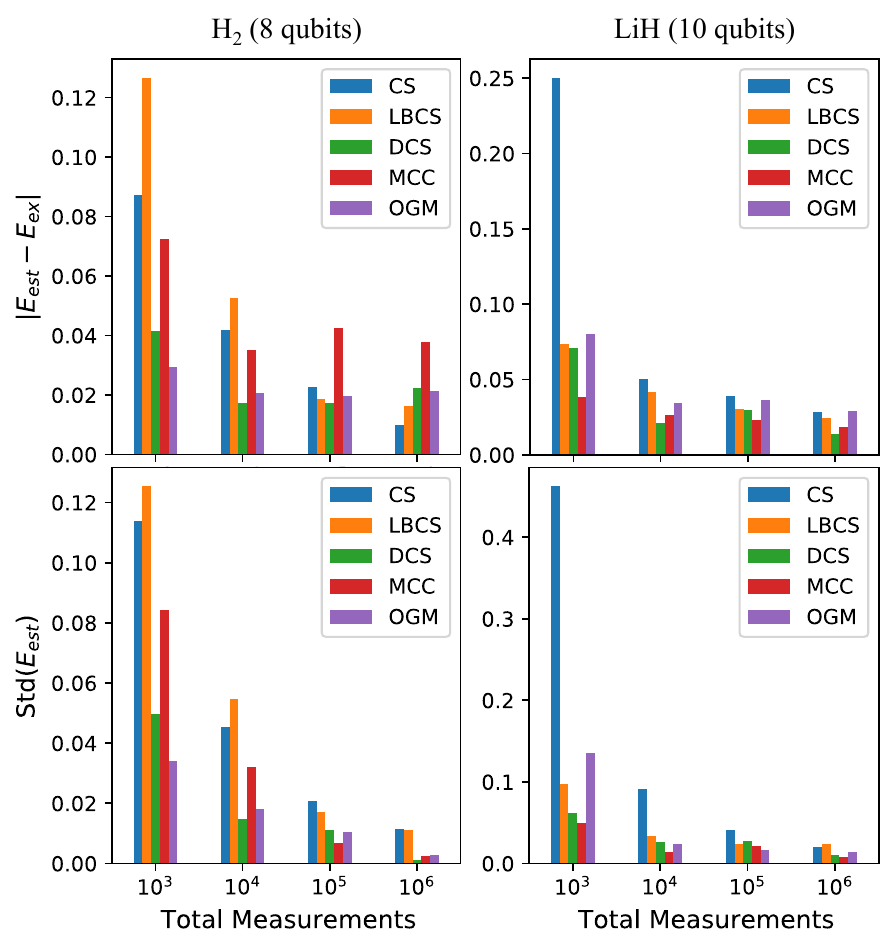}
    \vspace{-20pt}
    \end{center}
    \caption{\label{fig:qse_result}
    Mean absolute errors and standard deviations of estimated excited-state energies among various subroutines. 
     For detailed numerical values, refer to Appendix \ref{sec:fig_table}. 
    }
\end{figure}

\begin{table}[tbp]
    \centering
    \begin{tabular}{c|cc|cccc}
        \hline
        \multirow{2}{*}{Molecule} & \multicolumn{2}{c|}{$\#$ of Terms} & \multicolumn{4}{c}{Time of Preprocess (s)} \\
        & $H_d$ & $S_d$ & LBCS & DCS & MCC & OGM \\ \hline
        H$_2$, 4 qubits & 14 & 2 & 0.0150 & 0.3697 & 0.0003 & 0.0016 \\
        H$_2$, 8 qubits & 471 & 5 & 0.5413 & 0.0991 & 2.676 & 210.6 \\
        LiH, 10 qubits & 1936 & 10 & 5.897 & 11.36 & 1.583 & 14.74 \\ \hline
    \end{tabular}
        \caption{The number of Pauli terms in $H_d$ and $S_d$ for weakly correlated molecular Hamiltonians, along with the preprocessing time required for $H_d$.}
    \label{table:data_time}
\end{table}

In the context of measurement basis optimization for the Hamiltonian itself, it is known that there is a profound relationship between the time required for classical preprocessing and the estimation accuracy. However, whether similar properties generally hold in QSE is not trivial.
To investigate this, we have studied the relationship between the number of Pauli operators and the processing time required for each measurement optimization subroutine in Table \ref{table:data_time}. 
We find the general tendency that the MCC runs the fastest among the four, while the OGM is the most time-consuming.
Namely, there is a trade-off between accuracy and preprocessing time, which indicates that the choice of a measurement basis optimization subroutine depends on the classical computational resource.
For the sake of simplicity, in Sec.~\ref{subsec:convergence_adaptive_iteration}, we adopt the LBCS to demonstrate the effectiveness of the adaptive protocol.

We remark that, in the case of OGM, the measurement basis optimization is done faster for LiH (10 qubits) compared to H$_2$ (8 qubits). This is likely due to the symmetry of the Pauli operators of the Hamiltonian rather than the number of terms, which affects the optimization process dominantly.

\begin{figure*}[ht]
\begin{center}
     \hspace{-0.5cm}
    \includegraphics[width=43pc,angle=0]{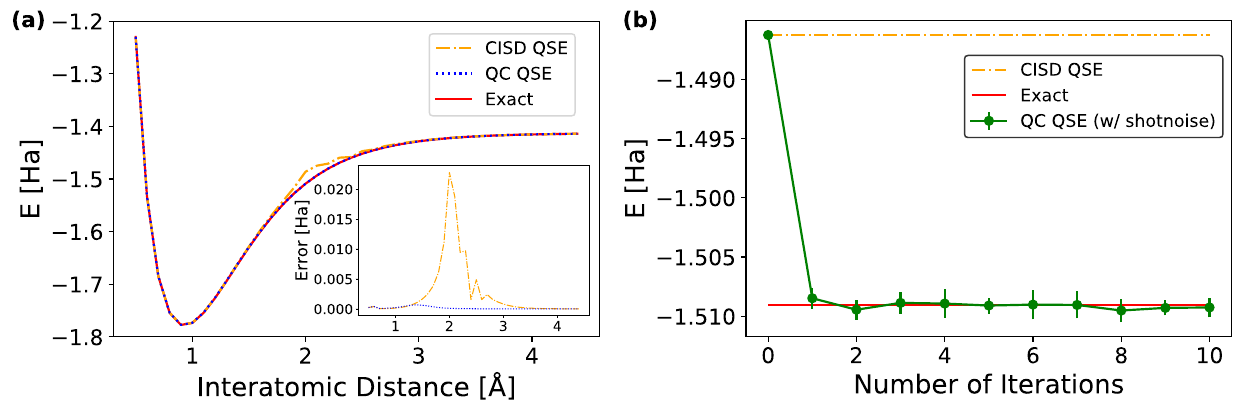}
    \caption{Benchmarking excited-state calculations for the H$_4$ molecule.
    The red solid lines represent the exact excited-state energies, while the orange and blue dashed lines depict the noiseless estimations using QSE with CISD and ground state (denoted QC), respectively.
    In both experiments, the regularization process was carried out with $n_{\rm lev}=20$ and  $\varepsilon=10^{-4}$ (see Appendix~\ref{sec:reg} for their definitions).
    (a) The excited-state energies with various interatomic distances. 
    The inset displays the absolute errors between the exact energies and the QSE results.
    (b) Demonstration of adaptive measurement strategy under shot noise with interatomic distance 2.0 Å.
    The green dots represent the averaged results over  10 trials of 10 adaptive iterations assuming $10^8$ shots at each iteration.
    Here we have employed the LBCS as the measurement optimization subroutine.
    }
    \label{fig:qse_2p1n}
\end{center}
\end{figure*}

\subsection{Convergenve of adaptive iteration}\label{subsec:convergence_adaptive_iteration}
Now we present a numerical demonstration of the adaptive scheme.
Here we consider the strongly correlated H$_4$ molecule using the STO-6G basis set ($N=8$ qubits), with the truncated Hilbert space spanned by $\mathcal{S}={\rm Span}\{a_i a^\dag_j a_k \ket{\psi}\}$ instead of $\mathcal{S}={\rm Span}\{ a_i \ket{\psi}\}$.

As is illustrated in Fig.~\ref{fig:qse_2p1n}(a),
the proposed method successfully converges to the exact energy at any interatomic distance, even in the strongly correlated regime of interatomic distance around 2.0 to 2.5 Å.
Using the regularization technique described in Appendix~\ref{sec:reg},
we find that the proposed method can stably simulate the excited-state energies within the chemical accuracy of 0.0016 Ha (see Fig.~\ref{fig:qse_2p1n}(b)).
It is worth mentioning that we have chosen the number of truncated eigenvalues in $\widetilde{S}$ to be $n_{\rm lev}=20$ in this numerical demonstration. As $n_{\rm lev}$ is increased, the differences due to the choice of the reference state, either CISD or the ground state, tend to decrease. 
In large-scale simulations where one resorts to Monte Carlo sampling on a computational basis, the statistical error from the sampling is one of the bottlenecks to achieve high accuracy~\cite{assaraf2017optimizing, choo2020}, and therefore we envision that we cannot take arbitrarily large $n_{\rm lev}$ so that a similar deviation occurs as well.

\section{Conclusion}
In this study, we have proposed an adaptive measurement strategy for situations where the classical description of the target observable explicitly relies on the unknown quantum state and is thus undecidable beforehand. 
We have in particular focused on measurement strategy for the quantum subspace expansion, and have proposed a scheme that initially determines the protocol based on classical simulatable states and then gradually updates based on measurement outcomes from the unknown quantum state. 
To validate the effectiveness of the proposed method, we have numerically simulated the excited-state energies of H$_2$, H$_4$,  and LiH molecules assuming that the exact ground states are realized on the quantum computer. 
The results show that, compared to a naive measurement strategy, the optimization on the measurement strategy exhibits up to a ten-fold reduction in the number of measurement shots, and further that the adaptive iteration converges rapidly even for the strongly correlated molecule of H$_4$.

Several future directions can be envisioned.
First, it is intriguing to derive the mathematical bound on the accuracy and iteration count of the adaptive updates based on information theoretic analysis.
This may lead to tighter bounds than what is currently known for the case with homogeneous measurement resource distribution~\cite{epperly2022theory}.
Second, it is practically important to investigate the performance of subspace methods in the context of quantum error mitigation~\cite{yoshioka_gse_2022}.
Quantum error mitigation techniques inevitably require exponential growth in the number of measurements~\cite{tsubouchi2022universal, takagi2022universal, quek2022exponentially}, while the exponent in the scaling heavily depends on the detail of the error mitigation methods.
Thus, it is crucial to seek how the adaptive strategy benefits the trade-off relation between bias and variance.

\section*{Acknowledgments}
This work is supported by PRESTO, JST, Grant No.\,JPMJPR2119, 
COI-NEXT program Grant No. JPMJPF2221, 
JST ERATO Grant No. JPMJER2302, 
and JST CREST Grant No. JPMJCR23I4, Japan 
and IBM Quantum. \\

\bibliographystyle{apsrev4-1}
\bibliography{qse_meas_strtgy}

\clearpage
\appendix

\section{Regularization of general eigenvalue problem} \label{sec:reg}
It is well-known that measurement shot-noise errors have a critical impact on the solutions of general eigenvalue problems. To address this issue, regularization of general eigenvalue problems has been formalized~\cite{epperly2022theory}.
Given the equation $\tilde H \vec{\alpha} = E \tilde S \vec{\alpha}$, we introduce parameters: a threshold $\varepsilon$ and a truncation number $n_{\rm lev}$ for the eigenvalues of the matrix $\tilde S$. 

Let $D$ and $V$ represent the eigenvalues and eigenvectors of the matrix $\tilde S$, respectively. An index set $I = \{i: D_i \geq \varepsilon\}$ generates the eigenvectors $V^{\geq \varepsilon} = V(:, I)$ whose eigenvalues are greater than or equal to the threshold $\varepsilon$. 
If the number of these eigenvectors is larger than $n_{\rm lev}$, we further truncate the indices by taking the largest $n_{\rm lev}$ of them to define the index set as $I'$. Then, the truncated eigenvectors are given by
\begin{eqnarray}
    V_{\rm reg} = V(:, I').
\end{eqnarray}
Using $V_{\rm reg}$, the matrices $\widetilde{H}$ and $\widetilde{S}$ are regularized as follows:
\begin{eqnarray}
\tilde H_{\rm reg} &= V_{\rm reg}^\dagger \tilde  H V_{\rm reg} \\
\tilde  S_{\rm reg} &= V_{\rm reg}^\dagger \tilde  S V_{\rm reg}
\end{eqnarray}
Solving the general eigenvalue problem $\tilde H_{\rm reg} \vec{\alpha} = E \tilde S_{\rm reg} \vec{\alpha}$ allows us to obtain the smallest eigenvalue as a solution to the initial general eigenvalue problem.

\section{Measurement basis optimization subroutines for explicit operators} \label{app:subroutine}
In this appendix, we provide a brief introduction to measurement strategies based on randomization and Pauli operator commutativity.
As measurement strategies based on randomized measurement, we introduce the Locally-biased Classical Shadow (LBCS) and Derandomized Classical Shadow (DCS). 
For commutativity-based measurement strategies, we then introduce Minimum Clique Cover (MCC) grouping and Overlapped Grouping Measurement (OGM).

\subsection{Methods based on random measurements}\label{subsubsec:lbcs}
Before introducing the LBCS and DCS, first we introduce the fully randomized measurement strategy which is based on the classical shadows (CS) of an unknown quantum state~\cite{huang_predicting_2020}.
The method can be summarized as follows: We define a set of measurement basis rotation unitaries denoted by $\{U\}$. Next, consider the averaged quantum measurement channel $\mathcal{M}(\cdot) = \mathbb{E}_{U}[\Pi(U\cdot U^\dag)]$ where  $\Pi$ is the projective measurement channel in the computational basis. 
We determine the distribution of $\{U\}$ so that the inverse channel $\mathcal{M}^{-1}$ is analytically given, often employing a uniform random distribution. This allows us to utilize the ``classical shadow" of the state $\hat{\rho} = \mathcal{M}^{-1}\left(U\ket{b}\bra{b}U^\dag \right)$ based on measurement outcomes $b$ to efficiently estimate various physical observables as
\begin{eqnarray}
    \ev{O} = \mathbb{E}_{U, b}[\hat{\rho} O].
\end{eqnarray}

When we employ only the Pauli measurements, the measurement strategy reduces to independently measuring the X, Y, and Z axes at each qubit, selected uniformly at random. 
In this case, the estimation variance grows exponentially with the locality of the observable, and thus the CS is not suitable for measuring physical observables related to systems such as quantum chemistry Hamiltonians transformed into qubit representation through the Jordan-Wigner transformation from the fermionic representation. In such cases, by selecting the set $\{U\}$ to follow a uniform distribution over the fermionic Gaussian unitary group or similar distributions, a polynomial overhead can be achieved~\cite{zhao_fermionic_2021}.

If one is interested in specific target observables, one may introduce bias into the distribution to improve the efficiency of the measurement strategy.
Among such techniques, here we introduce two notable ones: LBCS \cite{hadfield_measurements_2022} and DCS ~\cite{huang2021efficient}. 

The LBCS is very similar to CS, but differs in the way of determining a set of measurement basis rotation unitaries. For each qubit $i$, the probability of the $i$-th measurement basis being X, Y, or Z is all equally $1/3$ in CS for Pauli measurements, while in LBCS it is modified as $\beta_i(P_i)\ (P_i \in \lbrace X,Y,Z\rbrace)$ such that $\beta_i(X) + \beta_i(Y) + \beta_i(Z)=1$. The optimal bias $\beta_i$ is calculated by solving a convex optimization problem, based on the knowledge of the observable and a classical approximation of the quantum state.

The DCS is a method to improve efficiency by decisively determining the measurement basis~\cite{huang2021efficient}.
Concretely, one computes a cost function called the confidence bound that quantifies the probability of the estimator predicting the given observables with a given error.
It is argued that such a quantity is statistically sound, allowing the method to interpolate between completely randomized measurements and completely deterministic measurements.

\begin{table*}[thbp]
    \centering
    \renewcommand{\arraystretch}{1.3}
    \begin{tabular}{cc|c c c c c}
        \hline
        \multirow{2}{*}{Molecules} & Total & \multicolumn{5}{c}{Mean Absolute Errors(  $|E_{est} - E_{ex}|$ )}  \\
            & Measurements & \ \ \ \ CS \ \ \ \ & \ \ LBCS \ \ & \ \ \ DCS \ \ \ & \ \ \ MCC \ \ \ & \ \ \  OGM \ \ \ \\ \hline
        H$_2$, 4 qubits & $10^3$ & 0.0425 & $\leq 10^{-5}$ & $\leq 10^{-5}$ & $\leq 10^{-5}$ & $\leq 10^{-5}$ \\ \hline
        \multirow{4}{*}{H$_2$, 8 qubits} & $10^3$ & 0.0872 & 0.1267 & 0.0414 & 0.0726 & 0.0293 \\
            & $10^4$ & 0.0417 & 0.0526 & 0.0172 & 0.0352 & 0.0208\\
            & $10^5$ & 0.0228 & 0.0188 & 0.0174 & 0.0427 & 0.0197\\
            & $10^6$ & 0.0100 & 0.0164 & 0.0223 & 0.0377 & 0.0214\\ \hline
        \multirow{4}{*}{LiH, 10 qubits} & $10^3$ & 0.2503 & 0.0733 & 0.0711 & 0.0385 & 0.0800 \\
            & $10^4$ & 0.0501 & 0.0417 & 0.0213 & 0.0262 & 0.0342 \\
            & $10^5$ & 0.0389 & 0.0305 & 0.0297 & 0.0233 & 0.0361 \\
            & $10^6$ & 0.0282 & 0.0246 & 0.0135 & 0.0187 & 0.0294 \\ \hline    
    \end{tabular}
\end{table*}
\begin{table*}[thbp]
    \centering
    \hspace{-0.1cm}
    \renewcommand{\arraystretch}{1.3}
    \begin{tabular}{cc|c c c c c}
        \hline
        \multirow{2}{*}{Molecules} & Total & \multicolumn{5}{c}{Standard Deviations}  \\
            & Measurements & \ \ \ \ CS \ \ \ \ & \ \ LBCS \ \ & \ \ \ DCS \ \ \ & \ \ \ MCC \ \ \ & \ \ \  OGM \ \ \ \\ \hline
        H$_2$, 4 qubits & $10^3$ & 0.0516 & $\leq 10^{-5}$ & $\leq 10^{-5}$ & $\leq 10^{-5}$ & $\leq 10^{-5}$ \\ \hline
        \multirow{4}{*}{H$_2$, 8 qubits} & $10^3$ & 0.2765 & 0.2982 & 0.0703 & 0.0906 & 0.0315  \\
            & $10^4$ & 0.0818 & 0.0815 & 0.0387 & 0.0441 & 0.0194 \\
            & $10^5$ & 0.0262 & 0.0263 & 0.0153 & 0.0127 & 0.0162 \\ 
            & $10^6$ & 0.0113 & 0.0111 & 0.0013 & 0.0025 & 0.0028 \\ \hline
        \multirow{4}{*}{LiH, 10 qubits} & $10^3$ & 0.1377 & 0.1108 & 0.419  & 0.5687 & 0.1005 \\
            & $10^4$ & 0.0478 & 0.0616 & 0.0576 & 0.0293 & 0.0156 \\
            & $10^5$ & 0.0354 & 0.0439 & 0.0259 & 0.0196 & 0.0408 \\
            & $10^6$ & 0.0201 & 0.0232 & 0.0104 & 0.0079 & 0.0136 \\  \hline                                  
    \end{tabular}
    \renewcommand{\arraystretch}{1}
    \caption{Mean absolute errors and standard deviations in CS, LBCS, DCS, MCC and OGM (10 trials).}
    \label{table:data_compare_shadow}    
\end{table*}

\subsection{Methods based on commutativity}
\label{sec_OGM}
When we perform Pauli measurements, the observables must be qubit-wise commuting in order to be measured simultaneously. 
Namely, for $N$-qubit Pauli operators $Q = \bigotimes_{i=1}^N Q_i$ and $R = \bigotimes_{i=1}^N R_i~(Q_i, R_i \in \{I, X, Y, Z\})$, if it holds that $[Q_i, R_i]=0$ for all $i$, then $Q$ and $R$ can be simultaneously measured. Based on this fact, let us consider minimizing the number of measurements regarding an operator $H = \sum_i c_i P_i$ expressed as a sum of Pauli operators.
In other words, our goal is to identify the groups of Pauli operators $G_l = \{P_m^{(l)}\}_m$ that cover all the Pauli operators in the Hamiltonian as $\cap_{l=1}^{L}G_l = \{P_i\}_i$ under the condition that  Pauli operators within a group are simultaneously measurable. The assignment of groups is intended to minimize either the number of groups, represented as $L$, or the variance of the energy estimations.
While this problem can be formulated as the MCC problem when no overlap is allowed among measurement basis sets~\cite{verteletskyi_measurement_2020},
further efficiency can be achieved when overlap is allowed. In the following, we specifically introduce the OGM proposed by Wu et al. ~\cite{wu_overlappedgrouping_2023}.

The OGM is designed to minimize the cost function 
given by the probability distribution $\mathcal{K}$ of measurement bases ${\mu}$ for the operator $H$ as follows:
\begin{eqnarray}
loss(\mathcal{K})=\sum_{{P_i} \in \mathcal{S}} \frac{c_i^2}{\sum_{\mu: P_i \triangleright \mu} \mathcal{K}(\mu)}+\sum_{{P_i} \notin \mathcal{S}} c_i^2 T,
\end{eqnarray}
where $\mathcal{S}$ is a set of measurement bases, and  ${P_i} \in \mathcal{S}$ indicates that $P_i$ is measurable by one of the measurement bases $\mu$ in $\mathcal{S}$. Furthermore, $P\triangleright \mu$ indicates that the Pauli operator $P \in \{I, X, Y, Z\}^{\otimes N}$ can be diagonalized (i.e., measurable) through the measurement basis $\mu$. The second term represents a penalty imposed when $P_i$ is not included in any measurement set, where $T$ serves as a parameter determining the penalty scale. 
In all numerical demonstrations involving OGM, we used $T=1000$.
As discussed in Fig.~\ref{fig:qse_result} in the main text, we observe that the DCS and OGM yield steadily efficient measurements compared to CS, LBCS, or MCC, but the optimization procedure is more time-consuming. The choice between these strategies depends on the specific problem being addressed.

\section{Comprehensive Benchmarking Results for CS, LBCS, DCS, MCC and OGM} 
\label{sec:fig_table}
We present detailed numerical results in Table  \ref{table:data_compare_shadow} for the comparative benchmarking of CS, LBCS, DCS, MCC and OGM as shown in Fig.\ref{fig:qse_result}. 
In this experiment, we chose the regularization parameters $n_{\rm lev}$ individually such that the absolute error of estimation is minimized, and set $\varepsilon$ as the inverse square of the total measurements, acting as the threshold. 
The table also includes results for H$_2$ (4 qubits), which were not included in Fig.\ref{fig:qse_result}. Similar to TABLE \ref{table:data_protocol_advantage}, we conducted 10 trials and calculated the mean absolute errors and standard deviations for these results.
\end{document}